# Superconductivity with $T_c \approx$ 7 K under pressure for Cu-, and Au-doped BaFe$_2$As$_2$

Li Li,[1] David S. Parker,[1] Zheng Gai,[2] Huibo B. Cao,[3] Athena S. Sefat[1]

[1] Materials Science & Technology Division, Oak Ridge National Laboratory, Oak Ridge, TN 37831, USA
[2] Center for Nanophase Materials Sciences, Oak Ridge National Laboratory, Oak Ridge, TN 37831, USA
[3] Neutron Scattering Division, Oak Ridge National Laboratory, Oak Ridge, TN 37831, USA

**Abstract**

It is noteworthy that chemical substitution of BaFe$_2$As$_2$ (122) with the noble elements Cu and Au gives superconductivity with a maximum $T_c \approx$ 3 K, while Ag substitution (Ag-122) stays antiferromagnetic. For Ba(Fe$_{1-x}$TM$_x$)$_2$As$_2$, TM= Cu, Au, or Ag, and by doping an amount of x=0.04, *a*-lattice parameter slightly increases (0.4%) for all TM dopants, while *c*-lattice decreases (-0.2%) for TM=Cu, barely moves (0.05%) for Au, and increases (0.2%) for Ag. Despite the naive expectation that the noble elements of group 11 should affect the quantum properties of 122 similarly, they produce significant differences extending to the character of the ground state. For the Ag-122 crystal, evidence of only a filamentary superconductivity is noted with pressure. However, for Au and Cu doping (x≈0.03) we find a substantial improvement in the superconductivity, with $T_c$ increasing to 7 K and 7.5 K, respectively, under 20 kbar of pressure. As with the ambient pressure results, the identity of the dopant therefore has a substantial impact on the ground state properties. Density functional theory calculations corroborate these results and find evidence of strong electronic scattering for Au and Ag dopants, while Cu is comparatively less disruptive to the 122 electronic structure.

## 1. Introduction

High-temperature superconductivity (HTS) continues to puzzle the condensed-matter physics community. The antiferromagnetic undoped 'parents' of iron pnictide- and chalcogenide-based superconductors (FeSC) have small Fermi surfaces that are therefore sensitive to strain and small changes in chemical composition [1-4]. In fact, the application of pressure or the substitution of dopants in the parents of HTS can drive down their Néel ordering temperature ($T_N$) and yield HTS in them [5-8]. In the FeSC, there is a complex interplay and competition between multiple factors such as antiferromagnetism and superconductivity, the proximity of a lattice distortion to the $T_N$ and the associated nematicity [9-12], spin, charge, and orbital ordering [13-19], twinning [20, 21], and electronic and chemical disorder [22-26]. Therefore, although vast experimental and theoretical efforts have found and explained certain rich features in the FeSC, much remains unpredictable. For example, it is not yet possible to predict important features such as chemical doping trends, the reasons for the emergence of HTS with pressure or doping, or $T_c$ values. Hence, it is of interest to further delineate the behavior of these systems under apparently isoelectronic situations to isolate the effects of atomic size on these disparate properties.

As is well known, the undoped BaFe$_2$As$_2$ parent (known as '122') has a structural transition ($T_s$) and orders antiferromagnetically ($T_N$) below $T_s$=$T_N$≈135 K, with a stripe *C*-type spin arrangement in the Fe layers (where Fe spins are anti-parallel along the $a$ and $c$ axes, and parallel along the $b$ axis) [27-29]. For Ba(Fe$_{1-x}$TM$_x$)$_2$As$_2$ and by electron doping ($x$) of 122 using TM=Co [2, 30, 31], Ni [28, 32], Pd [33], Rh [33, 34], Ir [35], Cu [36], or Au [11], the $T_S$ and $T_N$ temperatures decrease and, intriguingly decouple ($T_S > T_N$) with increasing $x$, eventually giving superconductivity. In contrast, both transitions happen at the same temperatures for the hole-dopants TM=Cr [37, 38], V [12], Ag [39], Mn [40], and Mo [41], with no bulk superconductivity, while Ru doping creates electron carriers, with a $T_S = T_N$ overlap for small x, and superconductivity [42, 43]. Although it is unclear which dopants can favor HTS, several trends are apparent. For example, adding electron carriers seems to be necessary to induce superconductivity in transition-metal doped 122, as is suggested by these studies. Also, we have previously found that a decrease in the $c$-lattice parameter is needed for a substantial $T_c$ [44]. Moreover, it is noted that certain dopants (such as Co) introduce charge with minimal disturbance of the 122's electronic structure [39], while others, such as Ag or Cu appear more disruptive. Furthermore, local inhomogeneities in 122-doped crystals that are found in scanning tunneling microscopy/spectroscopy (STM/S) [45] are known to affect percolation and dictate the overall strength of superconductivity, while the applied pressure yields substantial changes to such local structural and electronic features [7,8].

In previous work, Cu- and Au-122 were found at ambient pressure to give maximum superconductivity at $T_c \approx 3$ K, at $x$ = 0.044 [36] and 0.031 [11], respectively, while Ag-122 remains antiferromagnetic up to $x$ = 0.045 doping level [39]. Given that properties, such as $T_c$ values, in this material class are known to be sensitive to lattice strain or pressure, here we study them by the application of external pressure. We indeed find a substantially higher $T_c$ ($\approx 7$ K) for Cu and Au doping, while Ag-122 exhibits an incipient, filamentary superconductivity, as in our previous work on CaFe$_2$As$_2$ [26].

## 2. Experiment

Single crystals of Ba(Fe$_{1-x}$TM$_x$)$_2$As$_2$ (TM = Cu, Ag, Au) (labeled as Cu-122, Ag-122, and Au-122, respectively) were grown out of FeAs self-flux technique [46]. To produce a range of TM dopant concentrations, small barium chunks, copper powder, silver powder, gold piece, and FeAs powder were combined according to various loading ratios in a glove box, and each placed in an alumina crucible. A second catch crucible containing quartz wool was placed on top of each growth crucible, and both were sealed inside a silica tube under ~1/3 atm argon gas. Each reaction was heated for ~24 h at 1180 °C, and then cooled at a rate of 1 to 2°C/h, followed by a decanting of the flux at ~1050 °C. The crystals were flat with dimensions of ~ 6 × 4 × 0.1 mm$^3$ or smaller. All crystals of TM-122 formed with the [001] direction perpendicular to the flat faces. The chemical composition of each crystal batch was measured with a Hitachi S3400 scanning electron microscope operating at 20 kV. Three spots (~ 80 μm) were checked and averaged on each random crystalline piece using energy-dispersive x-ray spectroscopy (EDS); the crystals had the same composition within each batch within error; no impurity phases or inclusions were detected. The samples are denoted by measured EDS $x$ values (each $x$ with a relative uncertainty of 5%) in Ba(Fe$_{1-x}$TM$_x$)$_2$As$_2$ (TM = Cu, Ag, Au) throughout this manuscript.

Bulk phase purity of Ba(Fe$_{1-x}$TM$_x$)$_2$As$_2$ crystals was checked by collecting data on an X'Pert PRO MPD X-ray powder diffractometer using monochromatic Cu $K_{\alpha 1}$ radiation in the 10-70° 2θ range, on ground crystals, each weighing ~ 30 mg collectively. Lattice parameters were refined from full-pattern refinements using X'Pert HighScore Plus software, at room temperature.

The STM/STS experiments were carried out with a mechanically sharpened Pt-Ir tip in an ultra-high vacuum variable temperature-STM chamber. The single crystal Ba-122 samples were mounted on a moly plate side by side to permit a direct comparison among the three samples by keeping the experimental conditions identical. The crystals were cleaved *in situ* at ~ 120 K and immediately transferred to the STM head which was precooled at 25 K. Topographic images were acquired in constant current mode with the bias voltage applied to the sample. Differential conductance (dI/dV) spectra were calculated numerically by taking the derivative of current-voltage (I-V) measurements.

Pressure experiments on electric transport were performed by using an HPC-33 pressure cell from Quantum Design (QD) Inc. Pressure was generated in a Teflon cup filled with Daphne Oil 7373 inserted into a Be-Cu pressure cell. The pressure was determined at low temperature by monitoring the shift in the superconducting transition temperature of pure lead. Resistivity measurements were performed using the QD Physical Property Measurement System (PPMS). A home-built uniaxial pressure cell based on a diamond presser was used to measure the magnetic susceptibility in the magnetic property measurement system (MPMS). The diamond cell applied pressure in the *c* direction while the sample was unconstrained in both the *a* and *b* directions.

## 3. Results and Discussion

The Bragg reflections were indexed using the tetragonal ThCr$_2$Si$_2$ tetragonal structure (*I4/mmm*), with no impurity phases. Fig. 1 plots normalized *a*- and *c*- lattice parameters as a function of *x* for Cu/Ag/Au doped Ba-122. For all the doping metals, the *a*-lattice parameter keeps increasing similarly with increasing *x*, with a change of +0.4% up to x=0.04. However, the *c* lattice parameter shows a contraction with Cu-doping (-0.2% for x=0.04), in contrast to the slight expansion with Au- (0.05% at x=0.04) and larger changes with Ag-doping of 122 (0.2% at x=0.04). Based on our previous finding [44], a decrease in the *c*-lattice parameter to be necessary for superconductivity, these results suggest that Cu-doping may be most prone to superconductivity [44].

Based on the bulk thermodynamic and transport measurement results on Ba(Fe$_{1-x}$TM$_x$)$_2$As$_2$ (TM = Cu, Ag, Au) crystals, temperature-composition (T-x) phase diagrams have been constructed for Cu-122 [36], Ag-122 [39] and Au-122 [11]. The change in $T_N$ versus *x* is summarized in Fig. 2: For Cu-122 and Au-122, $T_N$ is suppressed linearly with *x* and superconductivity with $T_C \approx$3 K is found in *x* = 0.044 and *x* = 0.031, respectively. A more detailed temperature-composition (T-x) phase diagram is shown for Ba(Fe$_{1-x}$Au$_x$)$_2$As$_2$ in the supplementary section of this manuscript, including more extensive neutron diffraction results on x= 0.012, 0.014, and 0.031 (compared to that of literature [11]). For Ag-122 and for *x* ≤ 0.019, $T_N$ follows the same trend as Cu-122 and Au-122, while for *x* > 0.019, it starts to diverge, approaching a potential saturated value (of $T_N$=78 K) around *x* = 0.045.

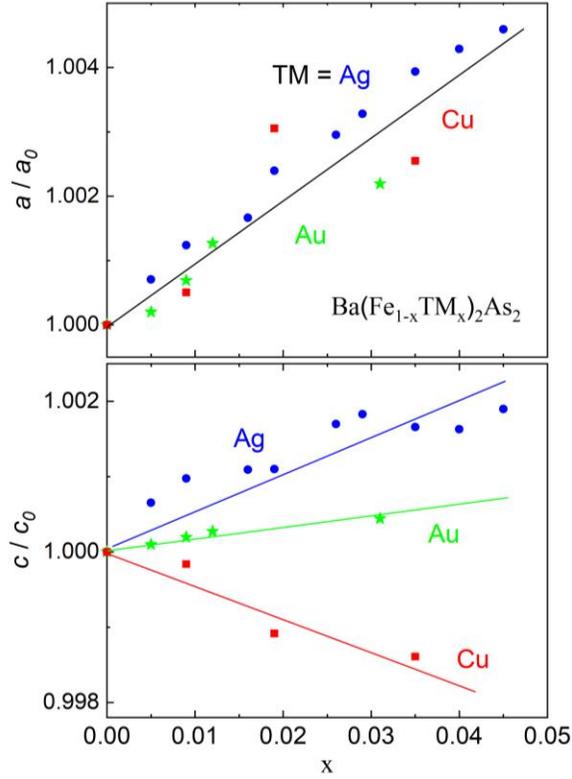

**Figure 1**. Room temperature lattice parameters of Ba(Fe$_{1-x}$TM$_x$)$_2$As$_2$ (TM = Cu, Ag, Au). The *a* and *c* parameters are normalized to the values of BaFe$_2$As$_2$ parent [$a_0$ = 3.9619(2) Å, $c_0$ = 13.0151(3) Å] as a function of TM concentration for each *x*.

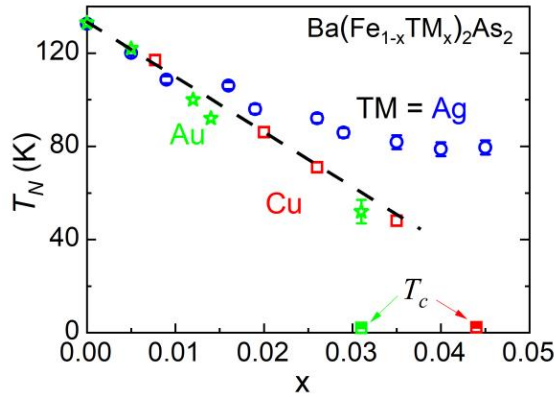

**Figure 2.** Néel temperature-composition ($T_N$-$x$) phase diagram for ambient pressure Ba(Fe$_{1-x}$TM$_x$)$_2$As$_2$ (TM = Cu, Ag, Au) following results reported in [11, 36, 39], with some new results added for Au-122 system (and detailed in supplemental document).

Next we study the effects of pressure on the bulk properties. Fig. 3(a) shows the temperature dependence of the resistance for Ag-122 with x = 0.005. With the application of pressure, the $T_N$ (inferred from the resistance) gradually decreases from 122 K at ambient pressure, to 115 K at 8 kbar, 110 K at 13 kbar, and finally down to 105 K at 20 kbar. In the meantime, a downturn in the resistance is observed at 22 K at 8 kbar, shifted to 24 K at 13 kbar, and 25K at 20 kbar. The resistance does not reach a zero value down to 2 K even at 20 kbar. However, as shown in the inset of Fig. 3(a), the downturn is shifted to lower temperature when applying a magnetic field at 8 Tesla, which typically indicates the existence of superconductivity. To confirm this likely filamentary superconducting feature, pressure study on magnetization is performed at low temperature below 30 K, here. Notably, a diamagnetic signal (Fig. 3b) is observed at low temperature when pressure is applied, and the transition temperature is close to that obtained from the resistance measurement at similar pressure (Fig 3a). This temperature of ~ 22 K is remarkably close to that of a similar transition which we recently observed in V-doped 122 [12] and suggests a possible common origin for these likely filamentary superconductivity behaviors.

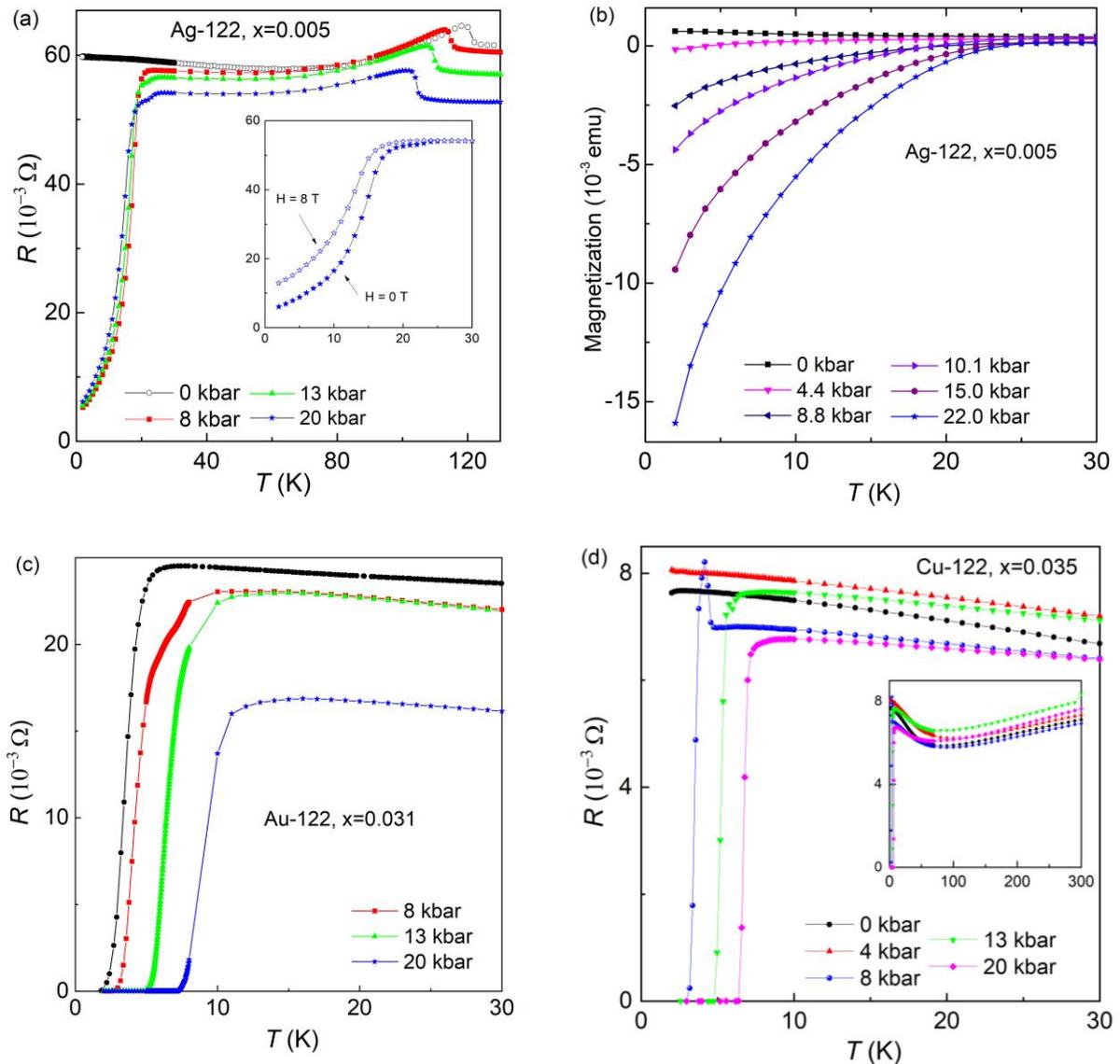

**Figure 3.** Temperature dependence of (a) resistance and (b) magnetic susceptibility for $Ba(Fe_{1-x}Ag_x)_2As_2$ (x = 0.005), (c) resistance for $Ba(Fe_{1-x}Au_x)_2As_2$ (x = 0.031), and (d) resistance for $Ba(Fe_{1-x}Cu_x)_2As_2$ (x = 0.035). Applied pressure levels are noted.

The temperature dependence of resistance for Au-122 with x = 0.031 is shown in Fig. 3(c). As previously noted, at ambient pressure, the resistance only shows a downturn without reaching a zero value at low temperature, which indicates filamentary superconductivity without a zero-resistance percolative network. However, with the application of pressure, we observe a robust bulk superconducting feature, with zero resistance appearing around 3 K at 8 kbar. This $T_C$ value increases with increasing pressure, reaching approximately 5 K at 13 kbar, and 7.5 K at 20 kbar. Similarly, for Cu-122 with x=0.035 (non-superconducting at ambient pressure), Fig. 3(d) shows that pressure of 8 kbar gives strong bulk superconductivity with $T_C$=3 K, which increases to ~ 5 K at 13 kbar, and $T_C$=7 K at 20 kbar.

Overall, these pressure results suggest that shrinking the lattice by the application of uniaxial pressure supports superconductivity. This is consistent with our previous finding [44] that, at ambient pressure only those dopants causing reductions in the c-axis parameter yield superconductivity. Here, however, it is the application of physical, not *chemical* pressure that is yielding the superconductivity, and these results are free of the confounding influence of charge doping. Hence pressure appears to be a particularly fundamental tuning knob for these phenomena, perhaps even more effective than charge doping itself. Electrostatic gating studies on these materials would therefore be of interest, as a means of isolating the effects of charge doping from the chemical pressure effects associated with atomic substitution.

The main question regarding the experimental results presented above which theory should attempt to answer is as follows: why do Cu-122 and Au-122 show superconductivity, with significant $T_c$ values as high as 7.5 K, while Ag-122 shows only a filamentary superconductivity, under pressure, and Co-122 shows $T_c$ values exceeding 20 K at ambient pressure?

A plausible hypothesis is that the dopants of Cu, Ag and Au present strong scattering centers that either preclude superconductivity or limit it to low $T_c$ values. Supporting this hypothesis are calculated plots (Fig. 5a) of the non-magnetic densities-of-states for 122 doped with each of these dopants, and also for cobalt. As we have argued previously [39], the DOS for these dopants are more alike than different, and all very different from that for Co. In the case of cobalt, the DOS effectively mirrors that of the entire 122 sample, which is suggestive of comparatively weak quasiparticle scattering [2]. It is well known that strong scattering tends to reduce $T_c$, with the details depending on the actual superconducting pairing symmetry and other factors, such as impurity concentration and disorder strength. Hence this plot is consistent with the experimental observation of a significant $T_c$ in Co-122. By contrast, for the three Cu, Ag, and Au dopants, the DOS attributable to the Cu group atom forms a separate, sharp impurity band, localized and in no way consonant with the overall DOS. This is suggestive of much stronger scattering.

STM and STS studies confirm the above theoretical results. STM imaging on all three crystals (Cu-, Ag- and Au-doped 122, with x=0.03) are very similar, showing coexistence of As and Ba terminated reconstructed areas on large flat terraces. Fig. 4 shows the comparison of the averaged local density of state (LDOS), (dI/dV)/(I/V), spectra over the whole scanning areas of 500 nm×500 nm, with 100×100 pixels collected at 25K. For all samples, the LDOS exhibits a V-shape around Fermi level, which is consistent with earlier reported tunneling spectroscopy measurements on the parent compounds of 122 Fe-based superconductors [47]. Consistent with the calculation, all three samples show similar behavior, especially in the empty states (positive bias). There exists slight discrepancy among the three spectrum, for example the Au-doped sample shows a little bit higher DOS in the filled states (negative sample bias), and the Au-doped sample and Ag-doped sample share a similar density of states peak at -0.1 V, but this peak is little bit weaker for the Cu-doped sample, because STM/STS is surface sensitive tool while our theoretical consideration does not include the breaking symmetry of the surface (reconstructions, surface relaxation and dopant inhomogeneity etc.)

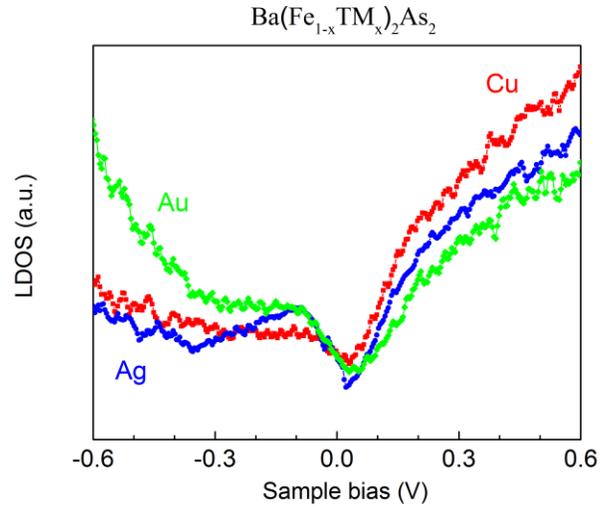

**Figure 4.** Direct comparison of the local density of states (LDOS) from $Ba(Fe_{1-x}TM_x)_2As_2$ with TM=Cu, Ag and Au doped samples (x = 0.03). The LDOS spectra were calculated from I/V curves averaged over 500 nm×500 nm areas from *in-situ* LT cleaved crystals, and the measurement was made at 25 K.

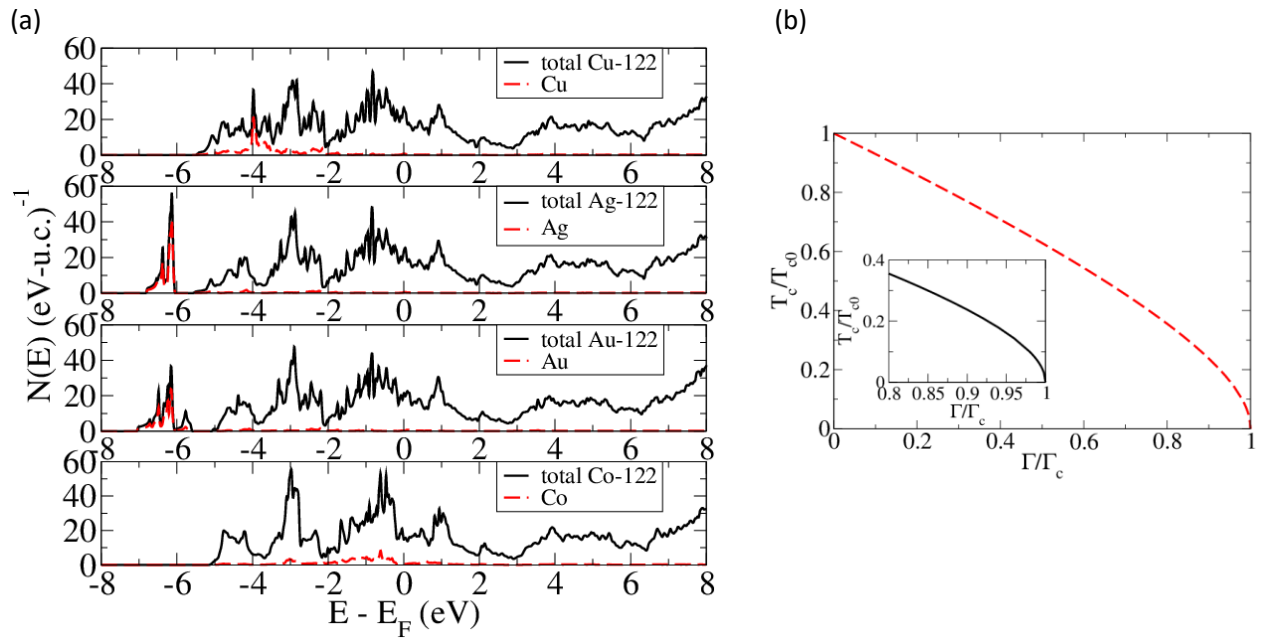

**Figure 5.** (a) The DOS of group 11 dopants compared with that for Co-doped 122. (2) The effect of impurity scattering rate on $T_c$.

However, this does not resolve the question of why Cu and Au dopants do exhibit bulk superconductivity under pressure, while Ag-122 does not. We note that this parallels the ambient-pressure situation. To address this question, we recall a seminal result from the early theory of superconductivity, the Abrikosov-Gor'kov [48] theory of the suppression of superconductivity by impurities. While this theory was originally constructed to apply to impurities in canonical *s*-wave phonon-mediated superconductors, with suitable modifications it can be applied to a broader class of materials, and for the foregoing qualitative discussion we therefore employ it. In Fig. 5b we show the suppression of $T_c$ by quasiparticle scattering. $T_c$ shows a steady reduction with increased scattering rate, and the slope of the $T_c$ curve increases greatly near the critical scattering rate $\Gamma_c$, where $T_c$ is suppressed entirely. We hypothesize that the group 11 dopants fall very near, but on opposite sides of, $\Gamma_c$. Since the $T_c$ curve slope is rather large at the critical concentration, the dopants to the left of this point show a non-negligible $T_c$ value, like 7 K, while the dopants to the right show no $T_c$. To illustrate, for a scattering rate only 5 percent short of the critical value, the critical temperature is already nearly 20 percent of the clean-limit value.

It is recognized that the above is only a rough qualitative explanation for the disparate $T_c$ behavior; in particular, it does not even address the effects of differing dopant concentration, lattice parameters, or for that matter, clustering and inhomogeneity. To make this more quantitative, we offer the following discussion. Any attempt to relate an overall scattering rate g due to a particular dopant, based on differences in the calculated densities-of-states, should obey several common-sense notions: it should depend in some monotonic manner on this difference; only states (i.e. occupied states) below the Fermi level should be considered (technically, since superconductivity may involve states slightly above $E_F$, this condition could be relaxed; we neglect this for simplicity); it should be positive-definite; there should be no region of energy giving negative contributions to G. With these conditions in mind, we postulate the following form for g. We stress that this form is not unique, but it has the virtue of simplicity and the quality of being readily calculable: $\gamma^2 = C \int_{E_0}^{E_F} (N_{tot}(E) - N_A(E))2 \, dE / (E_F - E_0)$.

Here $N_{tot}(E)$ is the total density-of-states, on a per transition metal atom basis, while $N_A(E)$ is the density-of-states from the dopant atom A, and the integral is taken from the bottom of the band to the Fermi level. We have normalized by the width of the band, which can vary depending on the dopant, and have used the squared difference to ensure positive definiteness. We take the square root based on the notion that the scattering rate should be proportional to some normalized measure of the difference in calculated DOS curves. C is presumed to be a constant for a given parent compound. For simplicity, for this discussion we choose C such that the largest scattering rate for our 4 dopants is unity. With these definitions in mind, we find the following values of g for the Ag, Au, Co and Cu-doped compounds: 1, 0.68, 0.32, and 0.64, respectively. We note immediately that the value of 0.32 for Co is much lower than the others, consistent with both one's intuition based on the similarity of the DOS curves for Co, and the observed substantial superconductivity for Co doping. The other three values are much larger, and we note that the largest value, unity for Ag corresponds well with both its particularly localized DOS in Fig. 5 and its complete absence of superconductivity, even under pressure, in the experiment. The values for Au and Cu are somewhat smaller and relatively near each other, consistent with their exhibiting superconductivity at comparable temperatures (although the particularly good agreement may be somewhat fortuitous). What the previous analysis suggests is that it may in fact be possible to make a theoretical estimate of the scattering strength attributable to a particular dopant, simply from a single first principles calculation. While this does not consider other superconductivity-relevant effects, such as the charge-doping induced modification of "nesting" features, it may yield insight about the potential of such dopants for inducing superconductivity.

## 4. Conclusions

An interesting observation is reported in our former paper, that when doping in Fe-plane of 122, only those which cause a decrease in $c$ (while maintaining a nearly constant $a$ near 4 Å) undergo a transition from antiferromagnetism to superconductivity [44]. One straightforward assumption is some way that can decrease the $c$-lattice parameters (shrinking the unit cell) may induce or enhance superconductivity in Cu-, Ag- or Au-122. Therefore, we performed pressure study on these. All crystals are prone to stronger superconductivity with pressure: Cu-122 and Au-122 crystals give higher $T_c \approx$ 7 K at 20 kbar, while Ag-122 crystal shows filamentary superconductivity. These results are consistent with a theoretical picture in which the scattering from all noble metal dopants (much greater than from cobalt) falls rather near a critical scattering rate at which superconductivity vanishes, where the dependence of $T_c$ on scattering rate is particularly strong.

**Acknowledgments**

The research is supported by the U.S. Department of Energy (DOE), Office of Science, Basic Energy Sciences (BES), Materials Sciences and Engineering Division. STM/S study was conducted at the Center for Nanophase Materials Sciences, which is a DOE Office of Science User Facility operated by Oak Ridge National Laboratory (ORNL). This research used resources at the High Flux Isotope Reactor, also a DOE Office of Science User Facility at ORNL.

# Supplemental Section

## Superconductivity with $T_c$= 7 K under pressure for Cu-, and Au-doped $BaFe_2As_2$

Li Li,[1] David S. Parker,[1] Zheng Gai,[2] Huibo B. Cao,[3] Athena S. Sefat[1]

[1] Materials Science & Technology Division, Oak Ridge National Laboratory, Oak Ridge, TN 37831, USA
[2] Center for Nanophase Materials Sciences, Oak Ridge National Laboratory, Oak Ridge, TN 37831, USA
[3] Neutron Scattering Division, Oak Ridge National Laboratory, Oak Ridge, TN 37831, USA

Single crystal neutron diffraction was performed using the four-circle diffractometer HB-3A at the High Flux Isotope Reactor (HFIR) at the Oak Ridge National Laboratory, to detect splitting between the structural ($T_S$) and Néel ($T_N$) transitions for $Ba(Fe_{1-x}Au_x)_2As_2$ with 1.2% ($x$=0.012), 1.4% ($x$=0.014), and 3.1% ($x$=0.031). Although we had presented a preliminary temperature-composition (T-$x$) phase diagram before [in ref. 11], we had only neutron diffraction on one doping level of 0.5% (x=0.005) and found $T_N$=122K and $T_S$=128 K; here we present a more accurate T-$x$ phase diagram by doing neutron diffraction on additional x.

The neutron wavelength of 1.542 Å was used from a bent perfect Si-220 monochromator. According to neutron diffraction data (Fig. S1), for x=0 and as found before, there is a simultaneous structural and magnetic transition. In the magnetic state, the spins are aligned along $a$-axis; the nearest-neighbor (nn) spins are antiparallel along $a$- and $c$-, and parallel along shortest $b$-axis. The nesting ordering wave vector is $q$ = $(101)_O$ or $(½ ½ 1)_T$, relative to the tetragonal (T) or orthorhombic (O) nuclear cells. The order parameter to the antiferromagnetic order is seen by the intensity of the magnetic reflection $(105)_O/(½½5)_T$; for tracking $T_S$, the intensity of the $(400)_O/220)_T$ nuclear peak was measured with warming. Similar to 122, the nuclear $(220)_T$ is expected to split to $(400)_O$ and $(040)_O$ orthorhombic Bragg reflections below $T_S$ in Au-122. The increased intensity of structural peak is due to reduced extinction effect by the structural transition from tetragonal to orthorhombic lattice. It is noticeable that $T_N$ and $T_S$ are identical for pure $BaFe_2As_2$ and split for doped samples. Both $T_N$ and $T_S$ gradually shift to low temperature with x increasing, which are summarized in the table in Fig. S1.

Based on the measurement results presented above, a more accurate T-$x$ phase diagram can be constructed [compared to ref. 11] for the $Ba(Fe_{1-x}Au_x)_2As_2$ system, shown in Fig. S2. Upon Au doping, the structural and magnetic transition temperatures decrease and the split between them rises. This Au-doped sample may contain a lot of chemical and electronic inhomogeneity, causing large splitting in the temperature difference of $T_S$ and $T_N$.

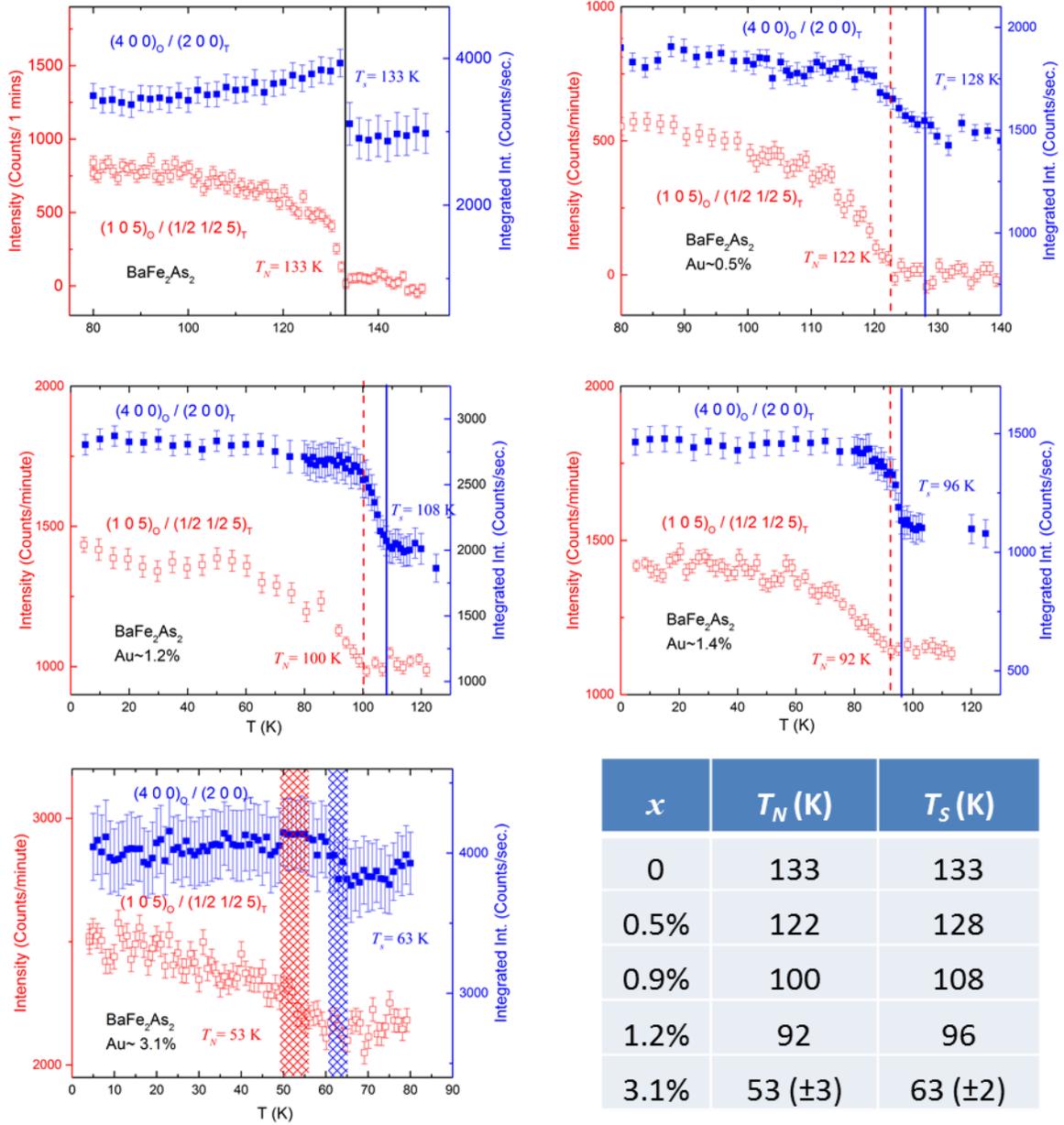

**Figure S1.** For Ba(Fe$_{1-x}$Au$_x$)$_2$As$_2$ crystals, neutron diffraction results for $x$ = 0, 0.5%, 1.2%, 1.4%, and 3.1%. The temperature dependence of Bragg reflection results upon warming. Integrated intensity of the nuclear peak (400)$_O$/(220)$_T$ and the peak intensity of the magnetic peak (105)$_O$/(½½5)$_T$ are represented by blue solid square and red empty square, respectively. The blue solid line marks the structural transition, and red dashed line marks the magnetic transition. The table summarizes $T_S$ and $T_N$ values.

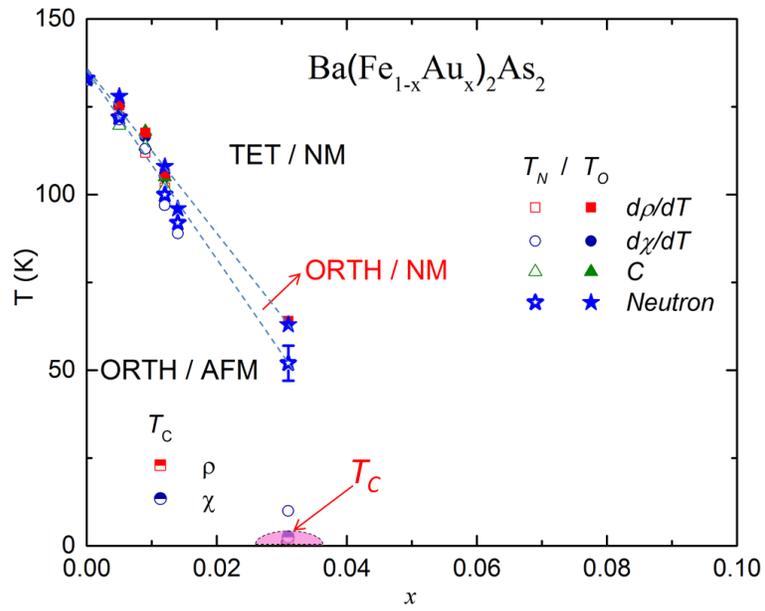

**Figure S2.** A more detailed and accurate T-*x* phase diagram for Ba(Fe$_{1-x}$Au$_x$)$_2$As$_2$ [adding to results from ref. 11].